\documentclass[
 reprint,
 amsmath,amssymb,
 aps,
]{revtex4-2}

\usepackage{graphicx}
\usepackage{dcolumn}
\usepackage{bm}

\begin{document}

\preprint{APS/123-QED}

\title{SALMON VR: Visualizing Light--Matter Dynamics}

\author{Tomohito Otobe$^{1}$}
 \email{otobe.tomohito@qst.go.jp}
 \author{Tatsuya Ogusu$^{2}$}
\author{Mizuki Tani$^{1}$}
\author{Shunsuke Yamada$^{1}$}
\affiliation{%
$^1$Kansai Insititute for Photon Science, QST,
            8-1-7, Umemidai, 
            Kizugawa,
            619-0215, 
            Kyoto,
            Japan
}%
\affiliation{%
$^2$Academeia Co., Ltd., 
178-4, Wakashiba,
Kashiwa, 
277-0871, 
Chiba,
Japan
}%

\date{\today}

\begin{abstract}
  This study presents SALMON VR, a visualization program 
  designed to visualize the time evolution of electronic density 
  changes and vector potentials in virtual reality (VR) space. 
  The time-series electronic density data computed by SALMON are 
  stored in CUBE format. SALMON VR processes these data to construct 
  isosurfaces of electronic density variations and two-dimensional 
  representations of vector potentials.
  Equipped with a user-friendly interface using VR technology, the program is 
  available in two versions: one for the Meta Quest 3 head-mount display (Meta Platforms Inc., California) and one for PCs. Atoms are displayed as spheres of different sizes and colors according to their elemental properties. This visual representation facilitates a deeper understanding of the complex interactions between light and electrons.
  Users can easily manipulate the isosurface values, speed of animation, and 
  color map of the vector potential. SALMON VR will enable researchers and educators 
  to enhance their understanding of physical phenomena and improve engagement 
  in learning environments.  
\end{abstract}

\maketitle

\section{Introduction}
In recent years, the visualization of data in computational physics and materials 
science has become an indispensable element in the advancement of research.
 Understanding complex three-dimensional data such as electron density variations, 
 current density, and vector potentials is crucial for gaining deep insights into physical 
 phenomena. In the last few decades, visualization technologies have evolved significantly, 
 particularly given the proliferation of virtual reality (VR) and augmented reality (AR) 
 technologies, which have laid the groundwork for exploring data from new perspectives \cite{VR_AR}.

Understanding the interaction between light and matter is critical for understanding the 
properties of materials.
In particular, the electron dynamics under coherent light pulses in
 the femto- and/or atto-second regions is a state-of-the-art topic in science \cite{Attophys}.
We have been developing a first-principles light--matter interaction simulator called Scalable Ab-initio Light--Matter simulator for Optics and Nanoscience (SALMON) \cite{SALMON}. 
SALMON enables first-principles calculations that couple electron dynamics with electromagnetic 
field dynamics. 
Our first principles approach on SALMON enables us to perform precise simulation of 
material optical properties not only in the linear regime but also in the non-perturbative regime. 
The data produced by SALMON include three-dimensional spatial data and their evolution over time. 
Therefore, these datasets are highly complex and multidimensional.

In this study, we have developed SALMON VR, which is an application for visualizing 
the data calculated by SALMON as continuous time-series files in 
CUBE format \cite{CUBEfile} in the VR space of the Meta Quest 3 (Meta Platforms Inc., California) and on a PC.  
SALMON VR offers users a real-time experience of the computational results, providing a novel platform enabling researchers and students to explore and analyze data on site.
Recent advances in visualization technologies enable users to explore large datasets in immersive environments. This dynamic use of SALMON's unique data has the potential to enhance understanding and facilitate new discoveries, allowing scientists to engage deeply with complex data through interactive experiences in virtual space.


\section{SALMON}
\subsection{Simulation with SALMON}
SALMON is an open-source program
for calculating the electron dynamics under a light field using time-dependent density functional theory (TDDFT) and Maxwell's equations in the time domain \cite{SALMON,Salmon:Online}.
The time-dependent Kohn-Sham (TDKS) equation \cite{TDDFT,SALMON} is used to calculate the electron dynamics, as follows:
\begin{equation}
\begin{split}i\hbar\frac{\partial}{\partial t}u_{b,{\bf k}}({\bf r},t)=\Big[\frac{1}{2m}{\left(-i\hbar\nabla+\hbar{\bf k}+\frac{e}{c}{\bf A}^{{\rm (t)}}(t)\right)}^{2}\\
-e\varphi({\bf r},t)+\hat{v}_{{\rm NL}}^{{{\bf k}+\frac{e}{\hbar c}{\bf A}^{{\rm (t)}}(t)}}+{v}_{{\rm xc}}({\bf r},t)\Big]u_{b,{\bf k}}({\bf r},t),
\end{split}
\label{1}
\end{equation}
where the scalar potential $\varphi({\bf r},t)$ includes the Hartree
potential from the electrons plus the local part of the ionic pseudopotentials,
and we have defined $\hat{v}_{{\rm NL}}^{{\bf k}}\equiv e^{-i{\bf k}\cdot{\bf r}}\hat{v}_{{\rm NL}}e^{i{\bf k}\cdot{\bf r}}$.
Here, $\hat{v}_{{\rm NL}}$ and ${v}_{{\rm xc}}({\bf r},t)$ are the
non-local part of the ionic pseudopotentials\cite{TM_PP} and exchange-correlation
potential, respectively. In addition to hybrid functionals, SALMON can use various other $v_{{\rm xc}}$.

The TDKS equation combined with Maxwell’s equations can be used to describe the
propagation of light.
SALMON has two types of light-matter coupling, multiscale and single-scale modes. 
Because SALMON VR is made for the single-scale mode \citep{SingleCell}, 
we focus on the single-scale mode in this paper.
Using Maxwell’s equations in single-scale mode, the propagation of the
electromagnetic fields in the form of the vector potential ${\bf A}({\bf r},t)$
is described as 
\begin{equation}
\left(\frac{1}{c^{2}}\ \frac{{\partial}^{2}}{{\partial t}^{2}}-\ \nabla^2\right){\bf A}\left({\bf r},t\right)+\frac{1}{c}\frac{\partial}{\partial t}\nabla\varphi({\bf r},t)=\ \frac{4\pi}{c}{\bf j}({\bf r},t),
\label{2}
\end{equation}
where ${\bf j}({\bf r},t)$ is the microscopic current density and $\varphi({\bf r},t)$ is the scalar potential.
Both the TDKS and Maxwell’s equations are discretized using uniform real-space grids.

\subsection{Output files}
SALMON has two options for writing the changes in electron density
\begin{equation}
    \delta\rho({\bf r},t)=\rho({\bf r}, t)-\rho({\bf r}, t=0),
    \end{equation}
    current density
    \begin{equation}
    {\bf j}({\bf r}, t)=-\frac{e}{m}\Re\sum^{occ}_{b,{\bf k}} u^*_{b,{\bf k}}\left(-i\hbar\nabla+\hbar{\bf k}+\frac{e}{c}{\bf A}({\bf r},t)\right) u_{b,{\bf k}},
\end{equation}
and vector potential (${\bf A}({\bf r},t)$). Here, $\rho({\bf r},t)$ is the electron density distribution at time $t$.
The first option, ``yn\_out\_dns\_ac\_je=y/n," outputs the electron density distribution, vector potential ${\bf j}$, and ionic coordinates in CUBE format \cite{CUBEfile} during the time propagation.
The data are output every time step, which is defined by the ``out\_dns\_ac\_je\_step" parameter.
The default output files are ``it \textit{time step number}\_drho.cube" for the electron density, ``it \textit{time step number}\_je\_x/y/z.cube" for the current density, and ``it \textit{time step number}\_Ac\_x/y/z.cube" for the vector potentials.
Here, ``\_x," ``\_y," and ``\_z" indicate the $x$, $y$, and $z$ components of the vector in Cartesian coordinates, respectively.
All CUBE files should be moved to the directory defined in the configuration file, which we discuss below.

\section{SALMON VR}
\begin{figure}
\includegraphics[width=80mm]{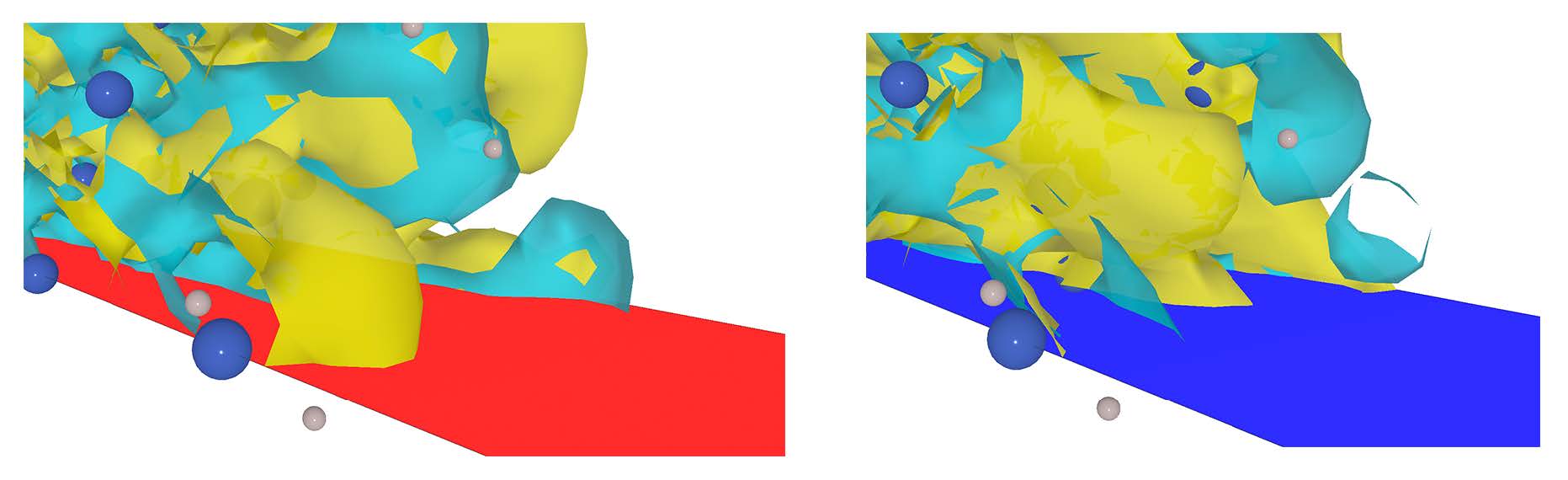}
\caption{Snapshot of an animation generated by SALMON VR: $\delta\rho$ under (left) positive and (right) negative vector potential values}
\label{fig1}
\end{figure}

SALMON VR \cite{SalmonVR:Online} is a visualization application 
that was developed using Unity \cite{Unity}.
In the default setting, SALMON VR
loads the CUBE files and displays the atoms $\delta\rho({\bf r},t)$ and ${\bf A}({\bf r},t)$.
As an example, we present a visualization of a silicon thin-film capped by hydrogen atoms.
Figure~\ref{fig1} shows a snapshot of the animation.
We used the local density approximation for the exchange correlation functional \cite{Perdew1981}.
The atoms are rendered as spherical objects, using different colors and sizes for each type. The color and size assignments follow the standard VESTA definition \cite{VESTA}. In Fig.~\ref{fig1}, 
the large blue spheres are silicon atoms and the small 
gray spheres are hydrogen atoms.
The values of $\delta \rho({\bf r},t)$ are rendered on two isosurfaces, where positive values are blue and negative values are yellow.
The isosurfaces are calculated 
using the marching cubes algorithm \cite{MarchingCubes}. 
The color range of an isosurface is automatically normalized by the maximum absolute $|\delta \rho(t)|$.

Vector potentials are rendered as mapped planes on the bottom boundary 
of the computational space. 
In many cases, the wavelength of the light is much larger than the target, and
density changes and vector potentials can be difficult to 
discern when they overlap.
To simplify visualization, the vertical component of 
the vector potential is averaged using an arithmetic mean. This process reduces visual noise and allows for a clearer understanding of the main trends.
The values of each grid are represented as red for positive values and blue
 for negative values. The closer the value is to zero, 
 the more transparent the point becomes, with zero being completely 
 transparent. The values are normalized so that the maximum 
 values are red (\#FF0000FF) and blue (\#0000FFFF).

Because the data are output in intervals defined by ``out\_dns\_ac\_je\_step," 
the animation does not run smoothly when the intervals are large.
Hence, SALMON VR uses the previous and next data to interpolate linearly as 
\begin{equation}
\rho({\bf r},t)=\frac{\rho_0({\bf r})(t_1-t)+\rho_1({\bf r})(t-t_0)}{t_1-t_0},    
\end{equation}
and generate images of the missing time steps.
Here, $\rho_{0(1)}(t)$ denotes the previous (next) loaded data, and $t_{0(1)}$ denotes the previous (next) time step. Moreover, $t_1$--$t_0$ are defined by the SALMON parameter ``out\_dns\_ac\_je." 
\subsection{Application for the Meta Quest 3}
There are two versions of SALMON VR, one is for a PC and the other is for the Meta Quest 3.
Using the Meta Quest 3 enables viewers to ``feel" the electron dynamics and vector potential in
VR space.
\begin{figure}
\includegraphics[width=80mm]{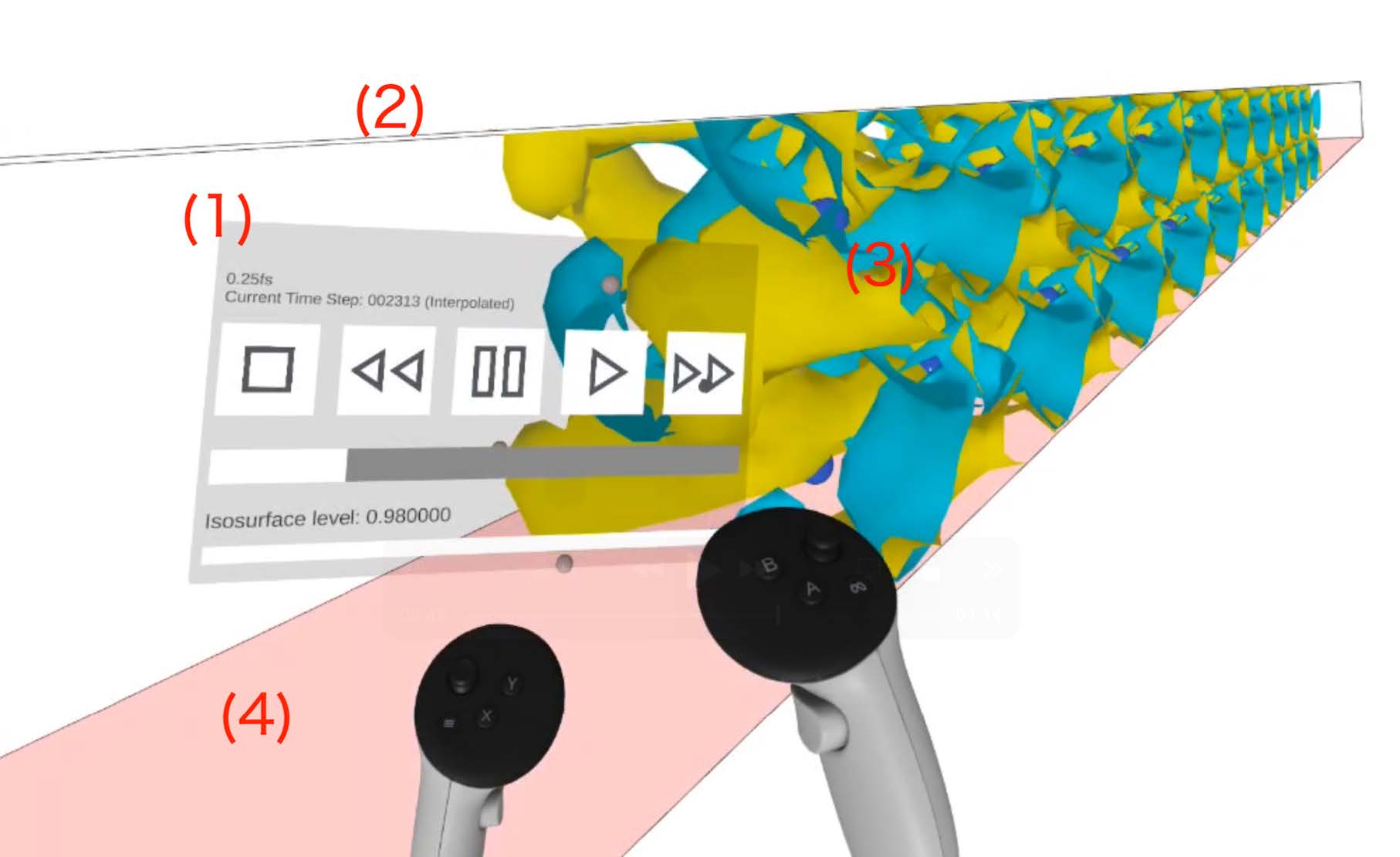}
\caption{Screenshot of SALMON VR on the . 
(1) GUI control panel for the left controller. 
(2) Computational box. 
(3) Isosurface and atoms. 
(4) Two-dimensional plot of the averaged vector potential on the bottom boundary of the computational box.}\label{fig2}
\end{figure}
Figure~\ref{fig2} presents a typical scene in the Meta Quest 3.
The graphical user interface (GUI) control panel  (Fig.~\ref{fig2}~(1)) appears above the left controller,
with the panel oriented to match the control surface where the stick and buttons 
are located.
The animation and isosurface level can be controlled using this 
control panel, as shown in Fig~\ref{fig3}.
The computational box is indicated by the black lines (Fig.~\ref{fig2}~(2)).
The atoms and $\delta \rho({\bf r},t)$ are shown 
in the computational box (Fig.~\ref{fig2}~(3)).
Finally, Figure~\ref{fig2}~(4) is the averaged vector potential.

The Meta Quest 3 has two controllers, one for the right hand and one for the left hand.
The viewpoint is controlled by these controllers.
Table~\ref{tab:meta_cont} lists the actions the user can take.
\begin{figure}
\centering
\includegraphics[width=70mm]{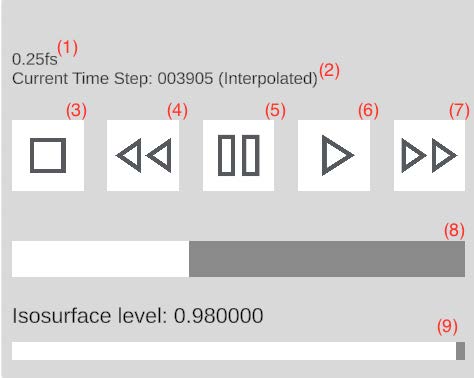}
\caption{Control panel on the Meta Quest 3. (1) Additional information, which can be changed in the configuration file. (2) Time steps. (3) Stop. (4) Double-speed reverse playback (the speed increases with every press). (5) Pause. (6) Play. (7)
Double-speed play (the speed increases with every press). (8) Seek Bar (the time can be manipulated by dragging). (9) Change the isosurface level (normalized to the maximum value)}\label{fig3}
\end{figure}
\begin{table}[]
    \centering
    \begin{tabular}{cc}
        {\bf Controller}&{\bf Action }\\ \hline
         Left Stick& 
         \begin{tabular}{c}
          Up/Down:\\
          Forward/Backward move\\
          Left/Right: \\
          Leftward/Rightward move\\           
         \end{tabular}
         \\  \hline
         Left Trigger& Speed up the movement\\ \hline
        
          Right Stick& 
         \begin{tabular}{c}
          Up/Down: \\
          Upward/Downward rotation \\
          Left/Right: \\
          Left/Right rotation 
         \end{tabular}
         \\  \hline
         Right Trigger& Speed up the rotation\\ \hline 
         
         A  & Play/Stop\\
         B  & Reverse/Stop\\
         Y  & Speed up \\
         X  & Speed up the reverse\\
    \end{tabular}
    \caption{Operation on the Meta Quest 3.}
    \label{tab:meta_cont}
\end{table}

\subsection{Application for MacOS and Windows}
\begin{figure}
\includegraphics[width=80mm]{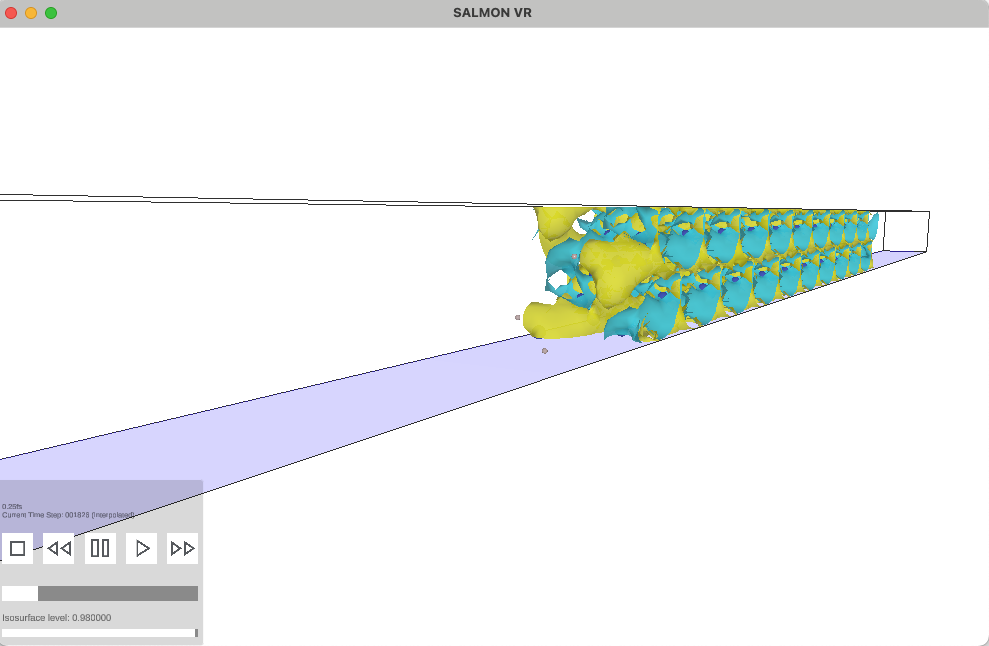}
\caption{Screenshot of SALMON VR on the PC. The control panel is on the bottom left.}\label{fig4}
\end{figure}

SALMON VR has also been implemented as a PC application (MacOS and Windows).
Although the Meta Quest 3 is a good device for interacting with the visualization, a typical PC monitor can be more suitable for sharing the image. 
Figure~\ref{fig4} indicates the panel of SALMON VR on MacOS.
With the exception of the control panel position, the definitions are the same as those of the Meta Quest 3.

The viewpoint and position can be controlled by the keyboard (Table~\ref{tab:Mac_cont}) and the mouse controller (Table~\ref{tab:Mac_cont2}).
\begin{table}[]
    \centering
    \begin{tabular}{cc}
        {\bf Key}&{\bf Action }\\ \hline
         w& Forward\\
         s& Backward\\
         a& Leftward\\
         d& Rightward \\
         e& Upward\\
         q& Downward\\
         ↑ & Upward rotation\\
         ↓ & Downward rotation\\
         ← & Left rotation\\
         → & Right rotation\\
    \end{tabular}
    \caption{Keyboard operation on the PC to change the viewpoint. The speed of movement and rotation are increased by the simultaneously pressing the Shift key.}
    \label{tab:Mac_cont}
\end{table}
\begin{table}[]
    \centering
    \begin{tabular}{cc}
        {\bf Operation}&{\bf Action }\\ \hline
         Scroll& Move forward/backward\\ \hline
         Drag with right button& Rotation\\ \hline 
         Drag with middle button& Move Up/Down/Left/Right\\
    \end{tabular}
    \caption{Mouse operation on the PC to change the viewpoint and position. }
    \label{tab:Mac_cont2}
\end{table}

\subsection{Configuration file}
\begin{table*}[th]
    \centering
    \begin{tabular}{cccc}
{\bf Parameter}& {\bf Value} & {\bf Type}  &{\bf Default}\\
\hline
CubeFilesDirectoryName     & Directory where the Cube files are located & Text 
&"Ac\_Je\_drho\_CUBE"\\
\hline
InfomationText & Additional text in GUI panel & Text & "0.25fs"\\
\hline
TimestepPerSecond & 
\begin{tabular}{c}
Specifies the speed of the animation. \\
The value specifies the time steps \\
per second \\
(for 100: 100 time steps per second).
\end{tabular}
& Integer & 100\\ \hline
DrhoFileFormat & File name for the isosurface& Text
& "*drho.cube"\\
\hline
AcFileFormat & File name for the 2D-plot on the floor & Text& "*Ac\_x.cube"\\ \hline 
EnableAcGammaCorrection & 
\begin{tabular}{c}
Gamma-corrected transparency \\
of the vector potential
\end{tabular}
& Boolean & false\\
\hline
AcGamma& 
\begin{tabular}{c}
Coefficient of gamma correction~\cite{GammaCorr} \\
(valid if "EnableAcGammaCorrection" \\ is true) 
\end{tabular}
& Real number& 2.2\\

\end{tabular}
    \caption{Parameters for the configuration file:config.json}
    \label{tab1}
\end{table*}

Although SALMON VR can be used without any preparation other than providing CUBE files,
we can change some parameters using the config.json configuration file.
An example of the config.json file is as follows:\\
\{\\
"CubeFilesDirectoryName": "Ac\_Je\_drho\_CUBE",\\
"InfomationText": "0.25fs",\\
"TimestepPerSecond": 100,\\
"DrhoFileFormat": "*drho.cube",\\
"AcFileFormat": "*Ac\_x.cube",\\
"EnableAcGammaCorrection": false,\\
"AcGamma": 2.2\\
\}.\\

Table~\ref{tab1} lists the definitions of the parameters for config.json.
``CubeFilesDirectoryName" defines the directory where the CUBE files to be loaded exist.
``InformationText" is the text displayed as the additional information.
``TimestepPerSecond" defines the speed of the animation (time steps per second). 
The default configuration describes the isosurface of $\delta \rho$ and the 2D plot of the x-component of ${\bf A}$.
By changing ``DrhoFileFormat" and ``AcFileFormat" we can create an animation of other data, such as ${\bf j}_e$.
The gamma corrected color map \cite{GammaCorr} can be used by setting ``EnableAcGammaCorrecti" to ``true" and selecting the correct coefficient of gamma correction if the ``AcFileFormat" color map is not suitable for the data.
Gamma correction is a non-linear operation to encode and decode transparency values. We employ the simple gamma correction
\begin{equation}
    f_{out}=f_{in}^{\gamma},
\end{equation}
where $f_{in}$ is the non-corrected data, $f_{out}$ is the corrected data, and $\gamma$ is the coefficient for gamma correction.
The new color scale is defined by $f_{out}$.
The config.json file is valid for both Meta Quest 3 and PC applications. 

\section{Conclusion}
 This paper presented SALMON VR, a versatile visualization tool with potential 
 for both research and education.  
 Its intuitive VR interface and interactive features overcome common challenges 
 in visualizing complex scientific data, thereby enhancing understanding and promoting 
 engagement among students and researchers alike.  
 The accessibility of SALMON VR, coupled with its potential for integration into 
 educational settings, has promise for improving the teaching and learning of optics 
 and laser physics.

 In addition to electron density and vector potential, 
 SALMON VR can be modified to display various time series 
 in CUBE files by changing the configuration file.
 Therefore, SALMON VR is not just an application for SALMON; it can visualize any
 4D data in VR space.
 \section*{\label{sec:acknowledgements} Acknowledgements}
This research was supported by MEXT Quantum Leap Flagship Program (MEXT Q-LEAP) Grant Number JPMXS0118067246 and KAKENHI Grant Number 24K01224. We thank Kimberly Moravec, PhD, from Edanz (https://jp.edanz.com/ac) for editing a draft of this manuscript. 


\bibliography{SALMONVR}


\end{document}